\documentclass[aps,tightenlines,11pt]{revtex4}
\usepackage{amssymb}

\def \beq {\begin{equation}}
\def \eeq {\end{equation}}
\def \lf  {\left (}
\def \rt  {\right )}
\def \pl  {\partial}

\begin{document}

\begin{titlepage}

\begin{center}

{\hbox to \hsize{\hfill CU-TP-1134}}

\bigskip

\vspace{3\baselineskip}

{\Large \bf The Volume of Black Holes}

\bigskip
\bigskip
\bigskip

{\large \sc Maulik~Parikh\\[3mm]}

{\it Department of Physics, Columbia University, New York, NY
  10027}\\[3mm]
{\tt mkp@phys.columbia.edu}

\vspace*{1.5cm}

\large{
{\bf Abstract}\\
}
\end{center}
\noindent 
We propose a definition of volume for stationary spacetimes.
The proposed volume is independent of
the choice of stationary time-slicing, and applies even though the
Killing vector may not be globally timelike. Moreover, it is constant in
time, as well as simple: the volume of a spherical
black hole in four dimensions turns out to be just ${4 \over 3} \pi r_+^3$.
We then consider whether it is possible to
construct spacetimes that have finite horizon area but infinite
volume, by sending the radius to infinity while making discrete
identifications to preserve the horizon area. 
We show that, in three or four dimensions, no such
solutions exist that are not inconsistent in some way. We discuss
the implications for the interpretation of the Bekenstein-Hawking
entropy.
\end{titlepage}

\newpage

\section{Introduction}

Time and space are often regarded as being interchanged across a black
hole horizon; the interior of a Schwarzschild black hole, for example,
can usefully be thought of as a collapsing universe. Moreover, what one
means by the volume of space depends on how spacetime is split into
space and time; spatial volume is not a slicing-invariant
quantity. Hence, at first sight, it does not seem to make any sense to
talk about the volume of a black hole.

This is unsatisfactory because one of the most celebrated facts about
quantum gravity is that the entropy is vastly reduced from what it would
have been in quantum field theory. The Bekenstein-Hawking entropy is
equal to $\rm{area}/4 {\it l}_p^2$, which, it is proclaimed, is
numerically much less than $\rm{volume}/{\it l}_p^3$. This begs the
question: what volume?

Now, questions of thermodynamics typically require thermal
equilibrium, and, geometrically, ``equilibrium'' means that the
spacetime possesses a symmetry under time translation i.e. there
exists a timelike Killing vector. Suppose we have such a timelike
Killing vector. Can one determine a volume in this more restricted
setting?

Remarkably, the answer is yes. In this paper, we will show that, if
the spacetime admits a Killing vector that is timelike in some region
then it is possible to define a thermodynamically meaningful notion of
volume. This is so even in the absence of a global timelike Killing vector;
indeed, the presence of a horizon implies that the Killing vector
becomes null there and, for nonextremal black holes, spacelike across
it. Nevertheless, the volume that we define is not only constant in time,
but also independent of the choice of stationary time slice (with the
one proviso that the asymptotic form of the metric also be
preserved).

Armed with a working definition of volume, an interesting next
question is: are there families of spacetimes whose horizons
have bounded area but whose volume can be arbitrarily large? For
example, one might try to send the mass of a black hole to infinity,
while simultaneously making discrete identifications on the
spacetime to preserve the horizon area. Were such a construction
to exist, it would be more than a curious fact: as we will argue, it
would strongly suggest that the Bekenstein-Hawking entropy counts
only the number of entangled states -- rather than the total number of
states -- in the quantum gravity Hilbert space. However, we will be
able to show that, at least in three or four spacetime dimensions, no
such families of spacetimes exist. We interpret this as evidence that the
Bekenstein-Hawking entropy might not be entanglement entropy.

\section{The Volume of a Black Hole}

For illustration, we will have in mind nonrotating black holes;
the final formula, though, requires only stationarity and applies
equally to rotating black holes. Consider then a spacetime with a
horizon and a line element of the form
\begin{equation}
ds^2 = - \alpha(r) dt_s^2 + {dr^2 \over \alpha(r)} + r^2 d
\Sigma^2_{D-2} (\vec{x}) \; ,
\end{equation}
where $d \Sigma^2_{D-2}$ can be taken to be the line element of a
maximally symmetric $D-2$-dimensional space. For instance, $\alpha$
could be
\begin{equation}
\alpha(r) = {2\Lambda \over (D-1)(D-2)} r^2 + \eta - {2 M \over
  r^{D-3}} \; . \label{alpha}
\end{equation}
One could also consider adding charge.
When $\Lambda \geq 0$, $\eta$ is $+1$ but, in AdS, $\eta$ can also be
$0$ or $-1$, corresponding to black holes with flat or negatively
curved horizons. The horizon is at $r_+$ where $r_+$ is the largest root of
$\alpha(r) = 0$. The time coordinate, $t_s$, is static time; the
metric is invariant not only under $t_s \to t_s + c$, but also under
$t_s \to -t_s$. However, the coordinate breaks down at the horizon,
as evidenced by the divergence of $g_{rr}$ and $g^{tt}$. To continue
through the horizon, one defines a new coordinate, $t$; static time
is then expressed as $t_s(t,r,\vec{x})$. In order for $\pl_t$ to
remain a Killing vector, the metric must be independent of $t$.
Writing
\begin{equation}
dt_s = {\pl t_s \over \pl t} (t,r,\vec{x}) dt + {\pl t_s \over \pl
r} (t,r,\vec{x}) dr + \vec{\nabla} t_s (t,r,\vec{x}) \cdot d \vec{x}
\; ,
\end{equation}
we see that $\pl_t$ is a Killing vector if and only if the
transformation takes the form
\begin{equation}
t_s = \lambda t +  f(r,\vec{x}) \; .  \label{transformation}
\end{equation}
Here $\lambda$ is a constant, which we can take to be positive to
preserve the orientation of time. In fact, our definition of volume will
require that $\lambda$ be restricted to 1:
\begin{equation}
t_s \equiv t +  f(r,\vec{x}) \; .  \label{timeslices}
\end{equation}
When $\eta \neq 0$, $\lambda$ can be set to 1 by demanding a fixed
asymptotic form of the metric. There remains an enormous class of
time-slicings, since each choice of the almost arbitrary function 
$f(r,\vec{x})$ defines a different time slice. To reduce clutter, we
will take $f(r,\vec{x})$ to be a function only of $r$. We then express the
line element in the new time coordinate as
\begin{equation}
ds^2 = - \alpha(r) dt^2 - 2 \alpha(r) f' dt ~ dr + dr^2 \lf {1 \over
  \alpha(r)} - \alpha(r) f'^2 \rt + r^2 d \Sigma^2_{D-2} (\vec{x}) \; .
\end{equation}
By choosing $f$ so that $f'$ is real and such that $g_{rr}$ stays
positive and finite, one obtains a stationary slicing that extends
through the horizon. Note that, although $\pl_t$ may become spacelike across
the horizon, the normal to a surface of constant $t$ is -- thanks to the
off-diagonal term -- everywhere timelike; such surfaces constitute
bona fide spatial sections. We would like to define an invariant
measure on these sections. First note that if we were to take the
volume to be the proper volume of a hypersurface of constant $t$, we
would not get an invariant volume because $g_{rr}$ manifestly depends on the
choice of time slice through its dependence on $f$. Indeed, by
considering a slicing that is nearly lightlike, one can arrange for
the proper three-volume to be as close to zero as one wants. If this
were the right notion of volume, it would not at all be clear that
holography entails a reduction in the degrees of freedom. 

Instead, observe that the determinant of the spacetime metric
\begin{equation}
- {\rm det}~g_{(D)} = + (\lambda^2)~r^{2(D-2)} ~{\rm det}~g(\Sigma) \; ,
\end{equation}
has no dependence on the time-slicing: $f'(r)$ drops out and $\lambda$
has been set to 1. This suggests the following definition of spatial
volume. Consider the differential {\em spacetime} volume
\begin{equation}
dV_D (t) = \int_t^{t+dt} dt' \int dr \int d^{D-2} x \sqrt{-g_{(D)}} \;
. \label{spacetimevolume}
\end{equation}
While the combination $d^D x \sqrt{-g_{(D)}}$ is
slicing-invariant (in fact, coordinate-invariant), $dV_D$ is not,
because the limits on the integral are defined in terms of a time
coordinate. However, if the time coordinate is of the form
(\ref{timeslices}) -- that is, if $\pl_t$ is a Killing vector -- then
the integrand is time-independent with time appearing in $dV_D$ only
through the multiplicative factor, $dt$.

We therefore propose that
\begin{equation}
V_{\rm space} \equiv {d V_D \over dt} \; .
\end{equation}
Equivalently,
\begin{equation}
V_{\rm space} \equiv \int d^{D-1} x \sqrt{-g_{(D)}} \;
. \label{volume}
\end{equation}
In other words, if, rather than using $\sqrt{g_{(D-1)}}$ as the
measure, one uses $\sqrt{-g_{(D)}}$ instead, then two things
happen. First, the volume is constant in time for all choices of
Killing time since the integrand is time-independent. And second, the
integral is invariant under stationary time slices.

Here is why. Imagine the spacetime integral, (\ref{spacetimevolume}),
as a Riemann sum of little strips, each of coordinate length $dt$,
lined up side-by-side from $r = 0$ to $r = r_+$.  According to
(\ref{timeslices}), a particular constant-time slice merely shifts
these strips up or down along the orbit of the Killing vector in an
$f$-dependent manner.  But the metric is unchanged under such
shifts. Hence the integral is invariant even though different
time-slicings correspond to integration over different spacetime
regions. After dividing out by $dt$, we therefore obtain an invariant
spatial volume. Nor is this construction affected by the nature --
timelike, spacelike, or null -- of the Killing vector. 

Actually, this is also the notion of volume that appears in
thermodynamics. To see this, write the partition function as
\begin{equation}
Z = \exp (-F \beta) = \exp \lf -\int d^D x \sqrt{-g_{(D)}} {\cal L}
\rt \; .
\end{equation}
Now the inverse temperature, $\beta$, is the period, $\int d \tau$,
where $\tau$ is a complexified time coordinate. Notice: there is no
$\sqrt{-g_{tt}}$ factor in $\beta$. Suppose the field is constant in
$\tau$. Then the free energy is
\begin{equation}
F = \int d^{D-1} x \sqrt{-g_{(D)}} {\cal L} \; .
\end{equation}
If the system is extensive, the free energy is proportional to the
volume. We see that this is not inconsistent with regarding $\int
d^{D-1} x \sqrt{-g_{(D)}}$ as the volume.

Let us now evaluate the volume for some simple spacetimes. For a
four-dimensional spherically symmetric black hole, we find that the
volume takes a satisfyingly familiar form:
\begin{equation}
V_{\rm spherical~hole} = \int_0^{r_+} dr \int_0^\pi d \theta \int_0^{2
  \pi} d \phi \sqrt{-{\rm det} g} = {4 \over 3} \pi r_+^3 \; .
\end{equation}
It is amusing that this is precisely the proper three-volume of flat
Euclidean space. A slicing in which the constant-time hypersurfaces
are flat is given by Painlev\'e coordinates, for which the line element
takes the form
\begin{equation}
ds^2 = -\alpha(r) dt^2 - 2 \sqrt{1 - \alpha(r)} dt ~ dr + dr^2 + r^2
d \Sigma_{D-2}^2 \; .
\end{equation}
These coordinates have already proven their utility in tunneling
calculations \cite{tunneling}.
But, more generally, such coordinates might not be defined globally
because the square root is required to remain real. So the volume
(\ref{volume}) should not be thought of as the proper volume of a
slice with $g_{rr} = 1$ as no such slice may exist.

The quantity $4 \pi r_+^3 / 3$ is precisely in accord with what an
observer in the time-independent region would consider to be the black
hole volume. To an internal observer,
however, $r$ is more like a time coordinate, and the result would
presumably not be interpreted as a volume. This is not surprising; it
is well-known that the thermodynamic properties of a black hole 
exist only from an outside point of view.

Finally, it should be clear that the above arguments did not rely on
the initial time coordinate, $t_s$, being static. Only stationarity --
the existence of a somewhere timelike Killing vector -- is
necessary. So the same volume formula applies to rotating black holes.
One finds that the volume of a four-dimensional Kerr black hole is
\begin{equation}
V_{\rm Kerr~hole} = {4 \over 3} \pi r_+ \lf r_+^2 + a^2 \rt \; .
\end{equation}

\section{Finite Area but Infinite Volume?}

As an elementary example, consider D-dimensional Rindler space, with $D
> 2$. In Cartesian coordinates, an observer moving with constant
acceleration in the positive $X^1$ direction, has a future Rindler horizon
described by the light-sheet $T = X^1$. The light-sheet has infinite
extent in the $X^i$ directions, for $i = 2 \ldots D-1$, so the horizon
has infinite area. The volume of the spacetime behind the Rindler
horizon is also intuively infinite. 
However, if we now make a toroidal compactification of all the
transverse directions,
\begin{equation}
X_i \sim X_i + L_i \; , \; \; \; i = 2 \ldots D-1
\end{equation}
the horizon area becomes finite: $A = \Pi_i L_i$. (The compactification
does not imply a dimensional reduction; the $L_i$ could be chosen to
be enormous compared with the $D$-dimensional Planck length.) However, because
$X^1$ is not identified, spatial sections behind the horizon are
noncompact and intuitively  have infinite volume. Thus this would
appear to be an example of a spacetime with a horizon of finite area
and infinite volume. However, it has been shown that Rindler space
with all but one spatial direction compactified is inconsistent
\cite{string1+1}. The formal proof of this claim consists of
demonstrating a contradiction between finite entropy and the
two-dimensional Poincare group, $ISO(1,1)$.

To find other spacetimes with this property, we note that Rindler
space is the infinite mass limit of a nonextremal black hole. Thus, in
general, what we would like to do is to take a spacetime with a
horizon, then send the radius of the horizon to infinity while making
discrete identifications to keep the area finite as the radius is sent
to infinity. More precisely, we would like to quotient by groups that
have the following properties:

(i) The group must be a subgroup of the isometry group of the
spatial section of the horizon. This is necessary so that the quotient
space has a well-defined metric.

(ii) The group must act freely on the spacetime. Otherwise, we would
introduce singularities. However,
we may allow a fixed point to occur at a point that is already singular
as singularities are not formally part of the manifold.

(iii) The fundamental domain must not have any cycles whose
length vanishes during the process of simultaneously blowing up the
horizon radius and quotienting by the groups. This is because, if there
were cycles of vanishing length, the gravity description could not be trusted;
winding modes of strings winding around the vanishing cycle would become
lighter than momentum modes. This is a restrictive requirement. It
implies that the identifications have to act
democratically in all dimensions along the horizon. Otherwise,
the directions in which they do act would be forced to
become vanishingly small to preserve the area as the radius is increased.

\subsection*{Spherical horizons}

Consider first spacetimes whose horizons, when sliced using stationary
time, are spheres. These have the isometry 
group $O(D-1)$. We need a family of discrete subgroups of arbitrarily
high order, so that, by quotienting with groups of ever larger order,
we can keep the area bounded even as the radius diverges.
For $D > 3$, there are two infinite families of
discrete subgroups of $O(D-1)$: the cyclic and the dihedral groups. The
cylic groups, $C_n$, have order $n$ and are isomorphic to $Z_n$.
They act by identifying points in the azimuthal direction: $\phi \sim
\phi + 2 \pi / n$. $C_n$ does not act freely because,
for example, it leaves the poles of the two-sphere fixed, in violation of
requirement (ii). The
dihedral groups, $D_n$, have order $2n$, and are isomorphic to $Z_2
\rtimes Z_n$. They are nonabelian and act freely. However, both $C_n$
and $D_n$ essentially act mainly along the azimuth. The
fundamental domain, after modding out by $D_n$, can be regarded as a
wedge extending down from the pole to the equator, much like a segment
of an orange. As the radius of the sphere becomes ever greater, the
width of the segment must vanish to preserve the area, thus violating
requirement (iii).

In three dimensions, stationary sections of the horizon are just
circles. So here we need subgroups of $O(2)$. Obviously, we can
mod out by $Z_n$. There are two spacetimes with horizons in three
dimensions: the BTZ black hole and three-dimensional de Sitter
space. For de Sitter space modding out by $Z_n$ in the angular
direction results in a conical singularity at $r = 0$. (After
appropriate relabelings, this can be regarded as a Schwarzschild-de
Sitter space, without an identification.)

Finally, consider the BTZ black hole with mass $M_0 > 0$. The line
element is 
\begin{equation}
ds^2 = -(r^2/l^2 - 8GM_0) dt^2 + {dr^2 \over r^2/l^2 - 8GM_0} + r^2 d
\phi^2 \; .
\end{equation}
The horizon is a circle which we can think of as a real line modded
out by Z. The entropy is just
\begin{equation}
S_0 = {\pi l \over 2 G} \sqrt{2 GM_0} \; .
\end{equation}
We now make the identification
\begin{equation}
\phi \sim \phi + {2 \pi \over n} \; ,
\end{equation}
which has the effect of changing the grading of the original
identification of the real line. Now let $M \to s M_0$ and define
$n = \sqrt{\lfloor s \rfloor}$. Then, as $s \to \infty$, $S \to S_0$. We see
that $M$ can be made arbitrarily large so long as $n$ is increased
suitably, without causing the entropy to diverge.

However, there is a problem here. Although discrete identifications
can be performed on the horizon, the uniqueness of the volume breaks
down because there is nothing to fix $\lambda$ to 1; time can be
rescaled. Indeed, after making the identification, if one defines the
new variables
\begin{equation}
\phi' \equiv n \phi \qquad r' \equiv r/n \qquad t' \equiv nt \; ,
\end{equation}
then the metric becomes that of a BTZ black hole with mass $M' =
M_0/n^2$, and no identification on the horizon. Thus area and volume
cannot be separately adjusted.

We have considered spherically-symmetric horizons, and shown that there are
no finite area and infinite volume solutions. We could also have tried
quotienting nonspherically-symmetric spacetimes such as the Kerr black
hole or Taub-NUT space. However, their isometry groups are just
subgroups of those of a sphere, and hence they also do not yield
finite area and infinite volume quotients.

\subsection*{Flat horizons}

Flat horizons exist in Rindler space, which we have already rejected,
and in AdS. The AdS black brane solutions have the line element
\begin{equation}
ds^2 = - \lf {r^2 \over l^2} -{2GM \over r^{D-3}} \rt dt^2 + {dr^2 \over {r^2
    \over l^2} -{2GM \over r^{D-3}}} + {r^2 \over l^2} \sum_{i=1}^{D-2}
    dx_i^2 \; . \label{blackbrane}
\end{equation}
The isometry group of the stationary slices is just $E(D-2)$ i.e.
$ISO(D-1)$. The lattice groups are discrete subgroups with no
fixed points. Thus we can make a toroidal identification on the horizon:
\begin{equation}
x_i \sim x_i + L \; .
\end{equation}
It is easy to see that this satisfies all the requisite
properties. After identification, the topology of the stationary
slices is now $T^{D-2}$. But since $\eta = 0$ in (\ref{alpha}),
$\lambda$ cannot be set to 1. So, again, just as with the BTZ black hole
(which can be regarded as a special case of (\ref{blackbrane})),
the volume cannot be invariantly defined.

\subsection*{Hyperbolic horizons}

AdS also has black hole solutions with hyperbolic horizons:
\begin{equation}
ds^2 = - \lf r^2/l^2 -1 -{2GM \over r^{D-3}} \rt dt^2 + {dr^2
    \over r^2/l^2 -1 -{2GM \over r^{D-3}}} + {r^2 \over l^2}
     d\Sigma_{D-2}^2 \; .
\end{equation}
Here $d \Sigma_{D-2}^2$ is the line element of a unit hyperbolic
space, $H^{D-2}$, a noncompact Riemannian manifold with constant unit
negative curvature (i.e. ``Euclidean'' anti-de Sitter space). The
isometry group of $H^{D-2}$ is $O(1,D-2)$, which is just the Lorentz
group.

Consider $D = 4$. Hawking's uniqueness theorem \cite{uniqueness}
on horizon topology does not apply to $AdS$ black holes; indeed, $H^2$
has infinitely many topologically inequivalent compactifications
\cite{pseudosphere}. One might hope that some of these might lead to
finite area and infinite volume spacetimes. However, the global
Gauss-Bonnet theorem says that the integral of the Ricci scalar is
related to the Euler characteristic, $\chi$, of the horizon:
\begin{equation}
{1\over 4 \pi} \int R d A = \chi = 2 - 2g \; .
\end{equation}
In two dimensions, compact oriented surfaces without boundaries or
punctures are topologically characterized by the genus, $g$. Thus we
find that the area is bounded from below:
\begin{equation}
A = 4 \pi (g - 1) r_+^2 \geq 4 \pi r_+^2 \; .
\end{equation}
We see that, irrespective of the compactification, the area becomes
infinite as the radius is sent to infinity.

In conclusion, we have shown that, in three or
four dimensions, there are no classes of spacetimes that have
bounded horizon area but unbounded volume. It would be interesting
to see whether this no-go theorem can be extended to higher
dimensions. Two loopholes in higher dimensions are that
asymptotically flat black geometries can have horizons with more
complicated topologies such as $S^1 \times S^2$ \cite{ring}, and
that higher-dimensional hyperbolic horizons are not subject to the
Gauss-Bonnet theorem. 

\section{Discussion}

We have seen that one cannot construct a family of spacetimes that
have horizons of bounded area but unbounded volume. It is intriguing
that in each case where this might have worked, something went
wrong: either there was a cycle of vanishing length (Schwarzschild
black holes), or there was a conical singularity (de Sitter space),
or the definition of volume became ambiguous (AdS branes and BTZ),
or there was a conflict with symmetries (Rindler space), or the area
itself diverged (hyperbolic horizons). Perhaps there is a deeper
reason why such a construction may be impossible.

One deeper reason might be a conflict with holography. The
statistical interpretation of Bekenstein-Hawking entropy, or
gravitational entropy, remains contentious \cite{trialogue}. One
school of thought holds that this entropy enumerates all possible
gravitational degrees of freedom within the volume enclosed by the
area. That is, it counts the total number of states in the quantum
gravity Hilbert space. An alternative interpretation is that
gravitational entropy counts only entangled states \cite{srednicki}.
The fact that the entropy scales as the area has two very different
implications from these two perspectives. The first implies that
quantum gravity is highly nonlocal, with far fewer degrees of
freedom than a local quantum field theory would have had. In
contrast, the second interpretation implies that quantum gravity is
local, locality being precisely the reason that the field deep
inside the hole is not entangled with the field outside.
Which of these two interpretations is correct is not immediately
obvious because we are unable to count the quantum gravity Hilbert
states directly. For example, in string theory, the counting of
microstates \cite{stromingervafa} is typically done in a dual picture
for which the string coupling is weak, leaving the gravitational
interpretation of the states unclear (though recent work attempts a
more direct approach \cite{mathur}).

Now, entanglement entropy is indifferent to the volume of
space. Indeed, its most appealing feature is that the entropy-area
relation appears quite naturally. From an entanglement entropy
perspective, there appears to be no reason why finite area but
infinite volume solutions should not exist. In fact, their existence
would have been evidence in support of entanglement entropy; the
alternate interpretation -- that the total number of Hilbert states in
an infinite volume is finite -- would then have seemed hard to
believe. However, the fact that no such solutions exist suggests --
assuming that the underlying reason is holographic -- that volume and
entropy are not independent. The Bekenstein-Hawking entropy may really
be counting all the quantum gravity Hilbert states.

\bigskip
\noindent
{\bf Acknowledgments}

\noindent The author would like to acknowledge useful conservations
with Robbert Dijkgraaf, Brian Greene, Daniel Kabat, Janna Levin, Juan
Maldacena, and Thanu Padmanabhan. M.~P. is supported in part by DOE grant
DF-FCO2-94ER40818 and by Columbia University's Frontiers of Science
program.

\end{document}